\documentclass{IEEEtran}
\usepackage{cite}
\usepackage{amsmath,amssymb,amsfonts}
\usepackage{graphicx}
\usepackage{textcomp,nicefrac}
\usepackage{lineno}
\usepackage[colorlinks=true,linkcolor=blue]{hyperref}
\usepackage{subfigure}
\usepackage{pdfpages}
\usepackage[normalem]{ulem}
\usepackage{siunitx}
\usepackage{hyperref}
\usepackage{subcaption}
\usepackage{caption}  
\usepackage{graphicx}
\usepackage{subcaption}
\usepackage{authblk}

\def\BibTeX{{\rm B\kern-.05em{\sc i\kern-.025em b}\kern-.08em
T\kern-.1667em\lower.7ex\hbox{E}\kern-.125emX}}
\begin{document}
\title{Development of Fast Front-End Electronics for the Muon Trigger Detector in the PSI muEDM Experiment}

\author{Tianqi Hu, Guan Ming Wong,  Chavdar Dutsov, Siew Yan Hoh, Kim Siang Khaw, Diego Alejandro Sanz Becerra, Philipp Schmidt-Wellenburg, Yuzhi Shang, and Yusuke Takeuchi
\thanks{This work is supported by the National Natural Science Foundation of China under Grant No. 120504102, China Scholarship Council No. 202206230093, and the Horizon EU program EURO-LABS. \textit{(Corresponding author: Kim Siang Khaw.)}}
\thanks{Tianqi Hu, Guan Ming Wong, Kim Siang Khaw, Siew Yan Hoh, Yuzhi Shang, and Yusuke Takeuchi are with the Tsung-Dao Lee Institute and School of Physics and Astronomy, Shanghai Jiao Tong University, 201210 Shanghai, China (e-mail: hutianqi@sjtu.edu.cn, wong.gm@sjtu.edu.cn, kimsiang84@sjtu.edu.cn, hohsiewyan@sjtu.edu.cn, yz.shang@sjtu.edu.cn, y.takeuchi@sjtu.edu.cn).}
\thanks{Diego Alejandro Sanz Becerra, Chavdar Dutsov, and Philipp Schmidt-Wellenburg are with the Paul Scherrer Institut, Forschungsstrasse 111, 5232 Villigen PSI, Switzerland (e-mail: diego.sanz-becerra@psi.ch, chavdar.dutsov@psi.ch, philipp.schmidt-wellenburg@psi.ch).}
}

\maketitle

\begin{abstract}
This paper outlines the design and development of a fast front-end electronic readout board for the muon trigger detector in the muEDM experiment at the Paul Scherrer Institute. The trigger detector, which consists of a gate and aperture detector, is strategically located at the end of the superconducting injection channel to generate trigger signals for a magnetic kicker, which activates upon the injection of muons into the central region of the storage solenoid. Within the magnetic field of the solenoid, the system configuration is optimized to meet stringent performance specifications, including minimal signal propagation delays, high detection efficiency, non-magnetic properties, and consistent operational stability under varying experimental conditions. Additionally, the design incorporates robust methods for rejecting positron contamination in the muon beamline. The results presented include performance evaluations from both bench testing and actual beam tests, highlighting the effectiveness and reliability of the electronic design.
\end{abstract}


\section{Introduction}
\label{sec:introduction}
\IEEEPARstart{T}{he} muEDM collaboration~\cite{Adelmann:2025nev, Adelmann:2021udj} at the Paul Scherrer Institute (PSI) aims to achieve a sensitivity of $6 \times 10^{-23}~\rm{e}\cdot\rm{cm}$ in the search for the muon electric dipole moment (EDM) using the "Frozen-Spin" technique~\cite{Farley:2003wt}. This represents an improvement of four orders of magnitude over the current best limit of $1.8 \times 10^{-19}~\rm{e}\cdot\rm{cm}$ (95\% Confidence Level) set by the Muon $g-2$ Collaboration at the Brookhaven National Laboratory~\cite{Muong-2:2008ebm}. Since the Standard Model prediction of the muon EDM is around $10^{-38}~\rm{e}\cdot\rm{cm}$~\cite{Pospelov:2013sca, Ghosh:2017uqq, Yamaguchi:2020eub}, any detected signal would clearly indicate physics beyond the Standard Model. Given that the EDM violates time-reversal (T) symmetry, it also violates Charge-Parity (CP) symmetry, assuming CPT symmetry holds. Consequently, the EDM is sensitive to new sources of CP-violation, potentially offering insights into the observed matter-antimatter asymmetry in the Universe~\cite{Sakharov:1967dj}.

\begin{figure}[htbp]
\centering 
\includegraphics[width=.50\textwidth, origin=c]{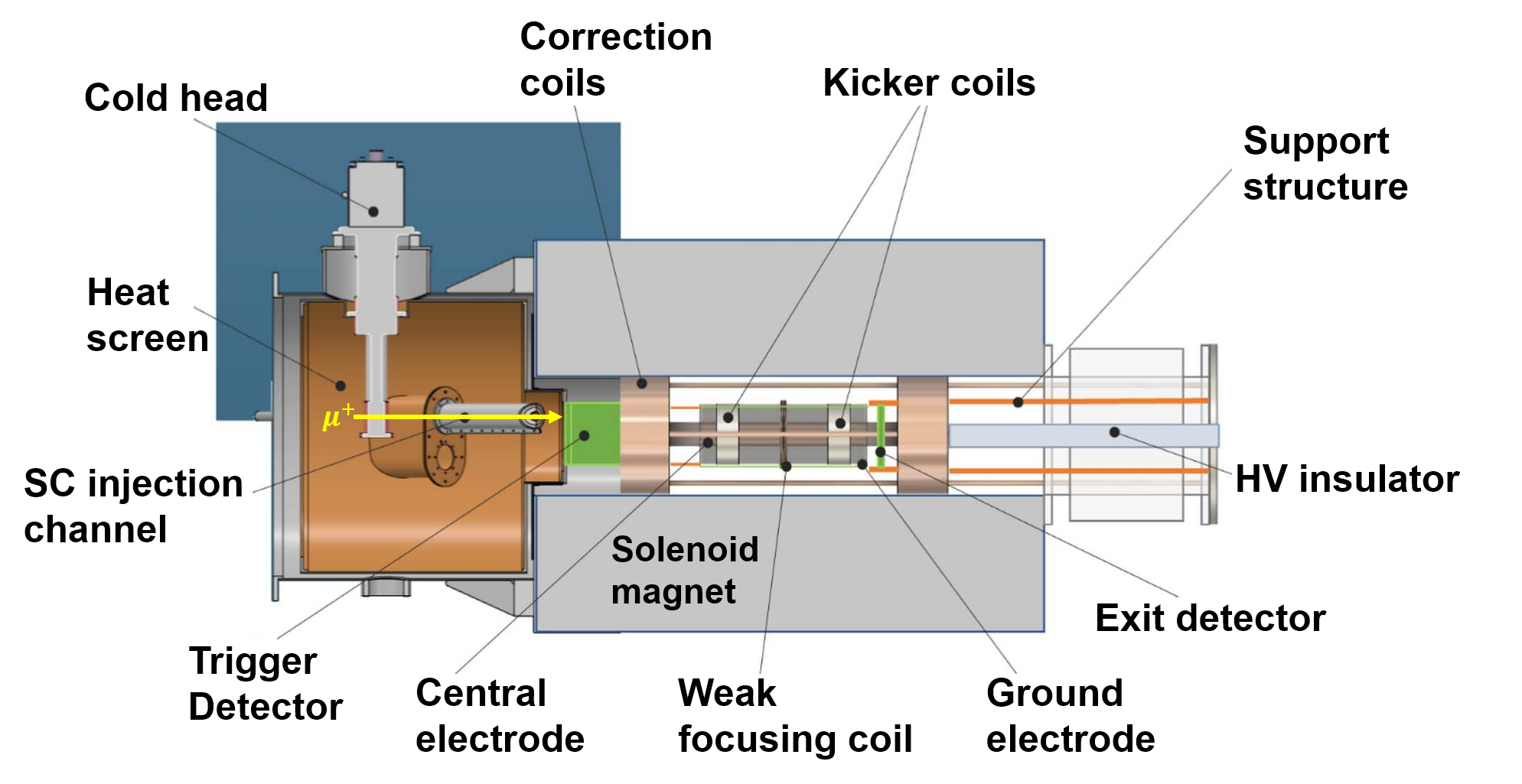}
\caption{Conceptual design of Phase I of the PSI muEDM experiment.}
\label{fig:PSImuEDM_ExperimentScheme}
\end{figure}

The muEDM experiment is divided into Phase-I and Phase-II, where Phase-I is a precursor experiment to demonstrate the frozen-spin technique. Phase-I's physics goal is to achieve a sensitivity of $4 \times 10^{-21}~\rm{e}\cdot\rm{cm}$. The concept design of the Phase-I experimental setup is illustrated in Fig.~\ref{fig:PSImuEDM_ExperimentScheme}. A polarized surface muon with a momentum of \SI{28}{MeV/c} is introduced into a solenoid featuring a \SI{3}{T} magnetic field through a superconducting injection channel~\cite{muEDM:2023mtc}, where it spirals into the central storage region. This injection method, known as Three-Dimensional Spiral Beam Injection, was proposed by the J-PARC Muon g-2/EDM Experiment~\cite{Oda:2021wgg}. Upon exiting this channel, a muon trigger detector detects the injected muon and sends a trigger signal to the kick coils—two circular coils carrying counter-propagating currents in an anti-Helmholtz configuration~\cite{muEDM:2024bri}. The kick coils generate a pulsed magnetic field that counteracts the muon’s longitudinal momentum, directing the muon into a stable storage orbit within a weakly focusing field created by the weak focusing coil. The activation of the kick coils is carefully synchronized with the muon’s transit through the weakly focusing field. Ultimately, the stored muon will decay into a positron, and the decay process depends on the muon’s spin orientation at the time of its decay. The measurement of the EDM signal will conclude by evaluating the asymmetry in the rates of upward and downward positrons exceeding a specified energy threshold, which varies over time.

\begin{figure*}[htbp]
\centering 
\includegraphics[width=\textwidth, origin=c]{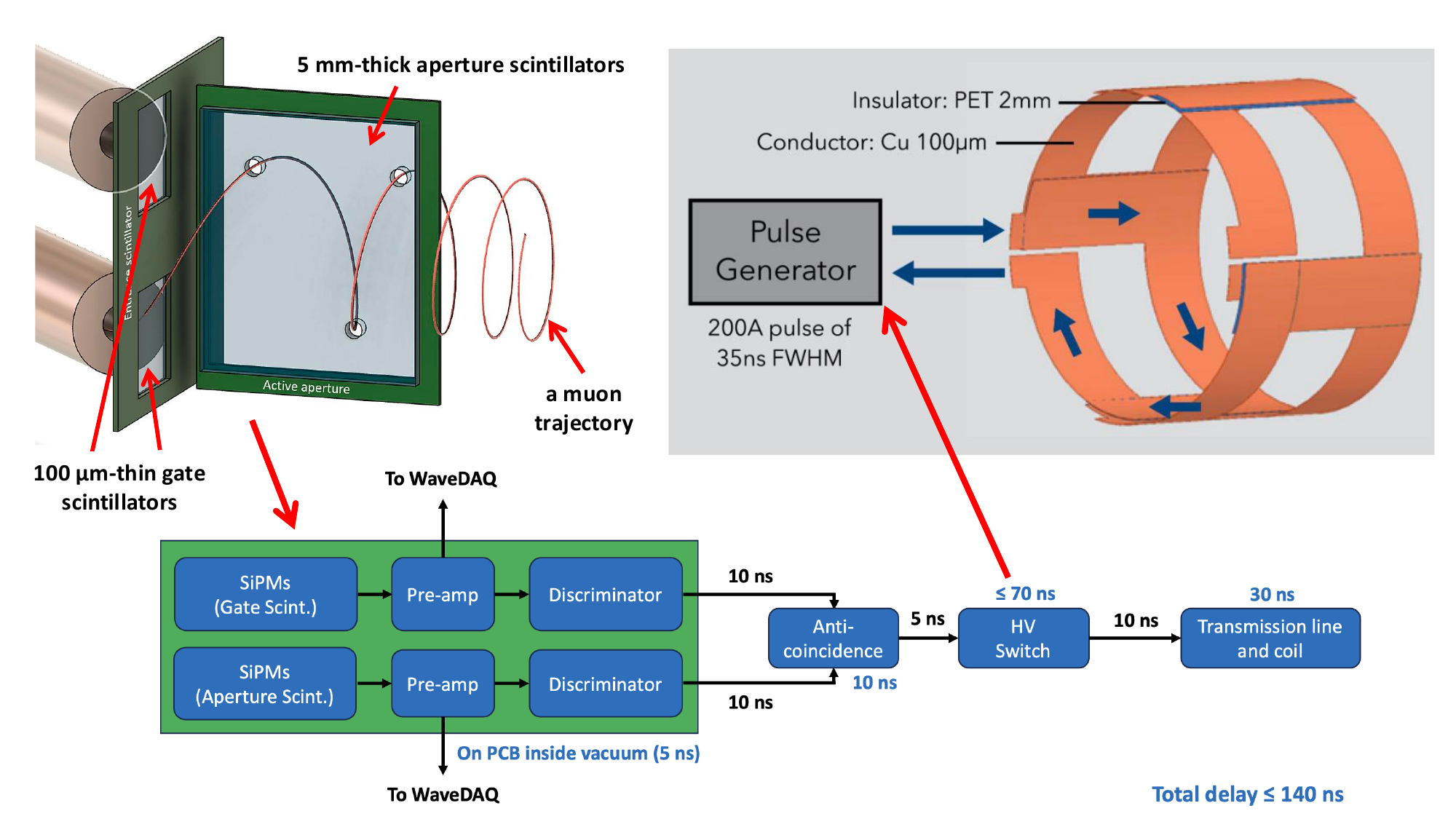}
\caption{A schematic layout of the electronic circuit responsible for delivering the current pulse to the kick coils is presented. The structure of the muon trigger detector is illustrated in the upper left corner, whereas the arrangement of the kick coils is depicted in the upper right corner.}
\label{fig:fast_electronics_timedelay}
\end{figure*}

The muon trigger detector represents a pivotal element of the experiment, composed of two distinct detectors: the \textit{gate detector} and the \textit{aperture detector}. Both detectors are precisely aligned with the nominal reference trajectory of the muons. The gate detector is responsible for identifying the injected muons and generating a trigger signal to activate the switch for the kicker coils. Conversely, the aperture detector inhibits the generation of trigger signals if the muons deviate from the acceptance phase space of the solenoid, thus creating an anti-coincidence system. This configuration effectively selects muons within the solenoid's acceptance, approximately $4 \times 10^{-3}$ of the incoming muons. It triggers the pulsed magnetic kicker required to deflect the muon into its designated storage orbit, thereby reducing the kicker power supply pulse rate from approximately 120\,kHz to 0.5\,kHz, which alleviates operational burdens and demands.

To meet the stringent demands of the muEDM experiment, several critical requirements must be fulfilled. First, the pulsed magnetic field must be activated within a latency range of 120\,ns to 150\,ns to ensure prompt muon deflection, which necessitates rigorous minimization of system-wide delays, especially in the muon trigger detector and the high-voltage switch of the pulse generator. Second, the front-end readout electronics of the muon trigger detector must operate reliably within a 3-T solenoidal magnetic field while maintaining minimal interference with the field. Additionally, the gate detector scintillator must be carefully optimized to minimize muon scattering while ensuring sufficient energy deposition and photon yield.

These requirements guided the design and development process. The scintillator thickness of the gate detector was optimized via Geant4~\cite{GEANT4:2002zbu} simulations to achieve a balance between scattering, energy deposition, and photon yield; the simulation results were validated experimentally, confirming the preservation of the beam’s phase space as it traverses the detector~\cite{Hu:2024avd}. In 2022, a proof-of-concept prototype using plastic scintillators read out by silicon photomultipliers (SiPMs) was constructed and tested under no-field conditions~\cite{Wong:2024vmo, Hu:2024avd}. The successful demonstration of its performance confirmed the feasibility of the selected technologies. Based on these results, a new detector system was subsequently developed to operate under Phase I experimental conditions.

This paper examines the research and development efforts behind the electronics of the muon trigger detector, focusing on its design, testing, and performance evaluation. The results demonstrate the detector's successful operation and highlight its capacity to meet the stringent requirements of the muEDM experiment.

\section{Technical requirements of the muon trigger detector}

The muon trigger detector system is designed to ensure precise and efficient selection of storable muons for the experiment. Its architecture and electronic design are optimized to meet stringent timing, environmental, and functional requirements under vacuum and high-magnetic-field conditions.

As illustrated in Fig.~\ref{fig:fast_electronics_timedelay}, the gate detector consists of two thin scintillator slices of \SI{0.1}{mm} thick. These are used for clockwise (CW) and counter-clockwise (CCW) beam injections. Combining data from both injection directions helps to significantly reduce systematic effects induced by the longitudinal electric field~\cite{Cavoto:2023xtw}. On the other hand, the aperture detector is made of a \SI{5}{mm}-thick scintillator with multiple holes aligned with the nominal muon trajectories. Muons falling within the solenoid’s acceptance region pass through these holes without interacting with the scintillator, distinguishing them from non-storable muons.

The electronic readout chain, also shown in Fig.~\ref{fig:fast_electronics_timedelay}, is carefully designed to maintain precise timing synchronization required for triggering the pulsed magnetic kicker to store the muons:

\begin{itemize}
    \item Signals from the gate and aperture scintillators are collected by Silicon Photomultipliers (SiPMs) and passed to the WaveDAQ system~\cite{Francesconi:2023cxt} for data recording.
    \item Simultaneously, fast trigger signals are generated via on-board discriminators with a latency of approximately \SI{5}{ns}.
    \item These signals are transmitted via coaxial cables and feedthroughs to a coincidence logic unit, introducing an additional \SI{10}{ns} delay.
    \item The logic unit processes these inputs and issues a trigger to activate the high-voltage (HV) switch for the kicker coils within another \SI{10}{ns}.
    \item Although the HV switch introduces an unavoidable internal delay, the upstream electronics are optimized to minimize cumulative latency.
    \item Once triggered, the current pulse travels through a vacuum feedthrough (adding approximately \SI{10}{ns}) and is conveyed to the solenoid center, where it must align precisely with the muon’s arrival for effective storage.
\end{itemize}

To meet the experimental demands, the following technical requirements must be satisfied:

\begin{itemize}
    \item \textbf{Latency and Speed:}
    \begin{itemize}
        \item The full signal propagation time from the detector to the anti-coincidence output must be below \SI{15}{ns}.
    \end{itemize}
    
    \item \textbf{Power and Circuitry:}
    \begin{itemize}
        \item The electronics operate on external supplies of $\pm$\SI{2.5}{V} and +\SI{3.3}{V}; SiPM bias voltage (\SI{32.5}{V}) is generated using an onboard DC-DC step-up converter.
        \item A signal splitter is integrated to simultaneously provide data for offline analysis and online discrimination.
        \item CW and CCW systems share a common readout circuit to reduce channel count and simplify layout.
    \end{itemize}
    
    \item \textbf{Logic and Integration:}
    \begin{itemize}
        \item An anti-coincidence circuit is embedded within the detector electronics to reduce delay from external cabling.
        \item All logic signals conform to LVTTL standards, ensuring compatibility with the kicker HV switch and DAQ system.
    \end{itemize}

    \item \textbf{Performance Metrics:}
    \begin{itemize}
        \item The system must reliably operate at a muon trigger rate exceeding \SI{120}{kHz}, with an expected stored muon rate of approximately \SI{0.5}{kHz}.
        \item Coincidence and anti-coincidence efficiencies must each exceed 95\%.
    \end{itemize}
\end{itemize}

\section{Design of the electronics for the muon trigger detector}

The schematic diagrams of the electronics for the muon trigger detector, illustrated in Fig.~\ref{fig:electronics_circuit}, comprise four components: the step-up DC-DC converter, gate electronics, aperture electronics, and Anti-coincidence electronics. The gate and aperture electronic circuits are similar, with the main difference being that the gate electronics connect to one SiPM, while the aperture electronics connect to six SiPMs in parallel with a single output. The gate Detector utilizes eight SiPMs, with four dedicated to CW readout and four to CCW readout, where both directions share a single set of electronics channels, resulting in four readout channels for the gate electronics. In contrast, the aperture detector requires only one aperture electronics channel.

The selected SiPM is the EQR20-11-3030S model from Novel Device Laboratory (NDL, Beijing), which features a peak photon detection efficiency (PDE) of approximately 46\% at \SI{420}{\nano\meter}, a gain of $8.2 \times 10^5$, a low dark count rate ($<$\SI{150}{\kilo\hertz\per\milli\meter\squared} at \SI{20}{\celsius}), and a low breakdown voltage of \SI{27.5}{\volt}, making it well-suited for timing-sensitive applications in our detector system. The choice was driven by its high PDE in the blue spectral range, compact active area of $3,\mathrm{mm} \times 6,\mathrm{mm}$, a large number of microcells (corresponding to a total sensitive area of \SI{2,500}{\milli\meter\squared}), and low operating voltage. The operating voltage used in this work was \SI{32.5}{\volt}, corresponding to an overvoltage of approximately \SI{5}{\volt}. The SiPMs are also anticipated to operate at a temperature of $25^\circ$C.

\subsection{Step-up DC/DC Converter}

\begin{figure}[htbp]
\centering 
\includegraphics[width=.48\textwidth, origin=c]{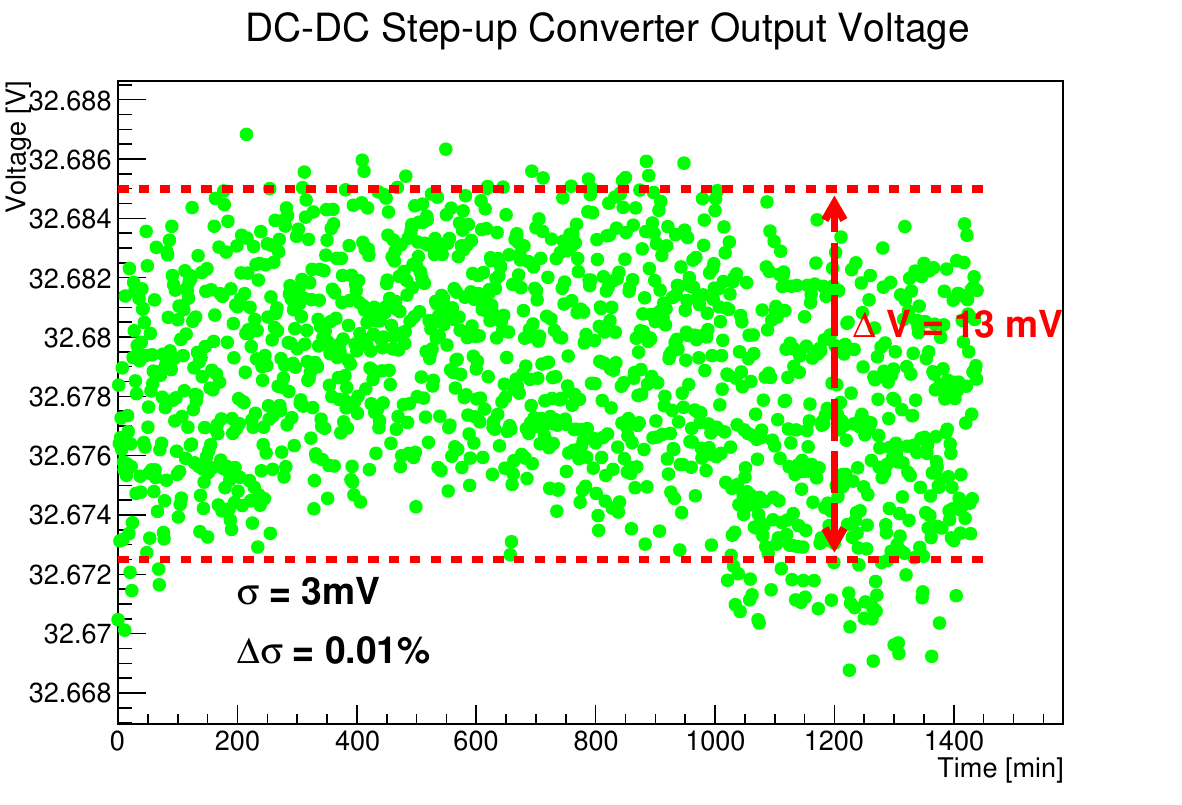}
\caption{The output voltage variation of the step-up DC-DC converter was evaluated at \SI{25}{\celsius}.}
\label{fig:DCDC_Output}
\end{figure}

\begin{figure*}[htbp]
\centering
\subfigure[The gate detector readout circuit]{\includegraphics[width=.95\textwidth,origin=c]{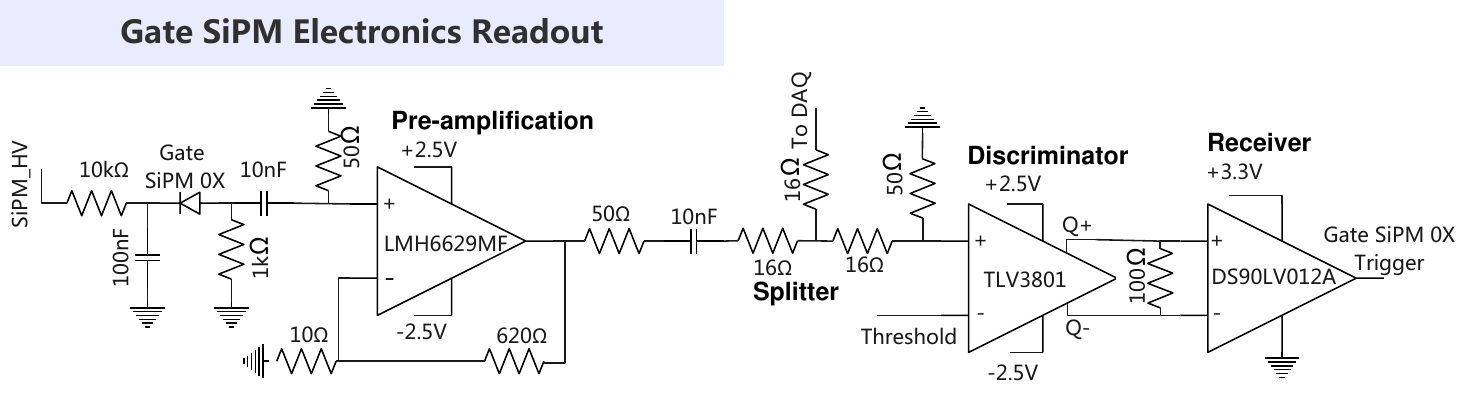} \label{fig:Gate_circuit}}
\qquad
\subfigure[The aperture detector readout circuit]{\includegraphics[width=.95\textwidth, origin=c]{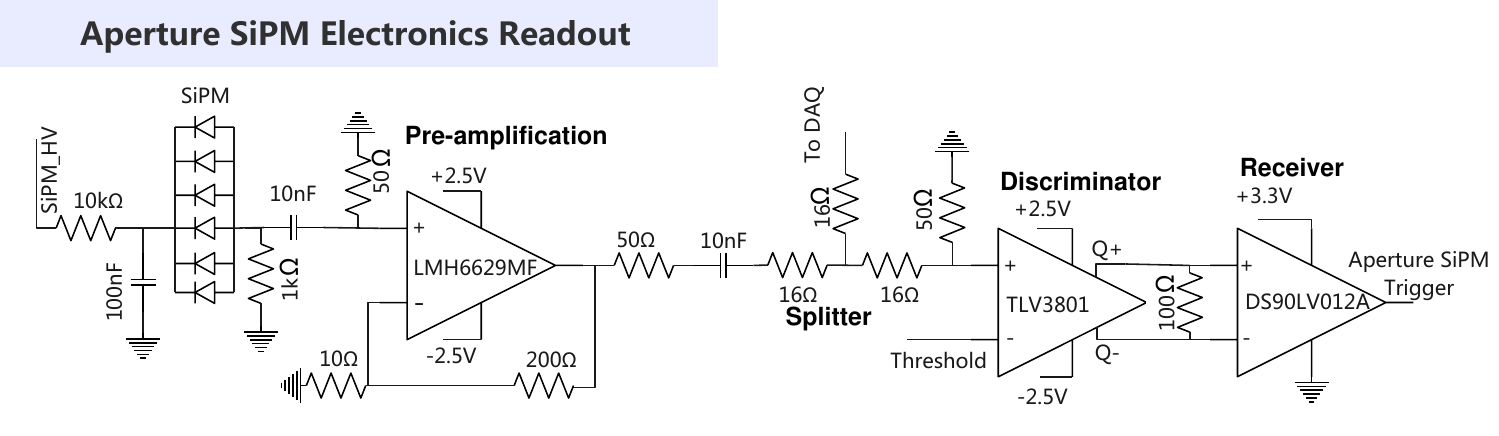} \label{fig:Aperture_circuit}}
\qquad
\subfigure[Electronics for the step-up DC-DC converter]{\includegraphics[width=.45\textwidth, origin=c]{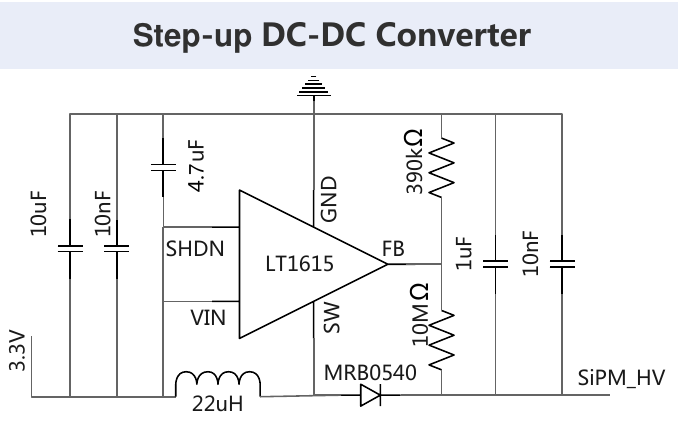} \label{fig:DCDC_Setup}}
\qquad
\subfigure[Anti-Coincidence Circuit]{\includegraphics[width=.48\textwidth, origin=c]{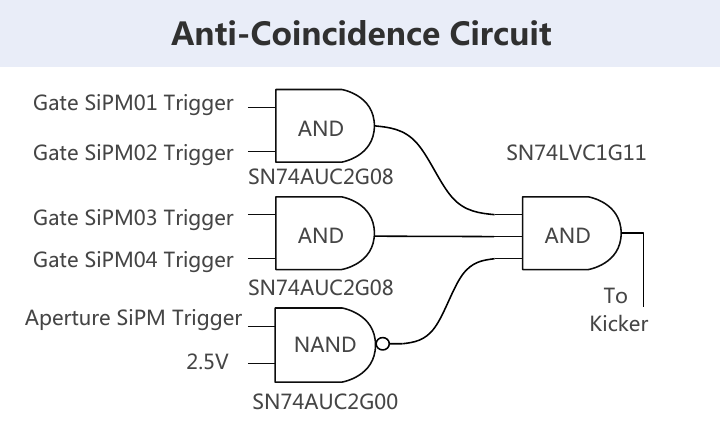} \label{fig:AntiCoincidence_circuit}}
\caption{\label{fig:electronics_circuit} Simplified schematic of the SiPM readout}
\end{figure*}

A step-up DC/DC converter was utilized to power the SiPMs, as illustrated in Fig.~\ref{fig:DCDC_Setup}. The chip used in the circuit is the Analog \href{https://www.analog.com/en/products/lt1615.html}{LT1615}. The driving voltage is +\SI{3.3}{V}. The output voltage variation over time is displayed in Fig.~\ref{fig:DCDC_Output}, demonstrating consistent stability in the output voltage. The average output voltage is approximately \SI{32.68}{V}, with a resolution of $\sigma=\SI{3}{mV}$. The fluctuation range is \SI{13}{mV}, which leads to a variation in the SiPM gain being $\sim 0.04\%$.

\subsection{Pre-amplifier Module and Splitter Module}

The amplifier chip used in the circuit is the Texas Instruments \href{https://www.ti.com/cn/lit/ds/symlink/lmh6629.pdf?ts=1724570048711&ref_url=https%253A%252F%252Fso.szlcsc.com%252F}{LMH6629MF/NOPB}, which has a bandwidth of \SI{900}{MHz} and a slew rate of \SI{1600}{V/ns}. The expected gain is calculated as \( \frac{620}{10} + 1 = 63 \) for the gate and \( \frac{200}{10} + 1 = 21 \) for the aperture electronics. A \SI{10}{nF} coupling capacitor is used as the filter.

A splitter circuit consisting of three $16~\Omega$ resistors, placed after the amplifier, is used to divide the amplified signal into two paths. One path directs the signal to the DAQ system for recording, while the other is sent to the discriminator. The splitter distributes the signal, with each path carrying a signal amplified 15.75 times compared to the original SiPM signal, after considering the impedance.

\subsection{Discriminator Module}

The discriminator module consists of two main components: the discriminator chip \href{https://www.ti.com/cn/lit/ds/symlink/tlv3801.pdf?ts=1724579913716&ref_url=https%253A%252F%252Fitem.szlcsc.com%252F}{TLV3801DSGT} and the Low Voltage Differential Signal (LVDS) receiver chip \href{https://www.ti.com/cn/lit/ds/symlink/ds90lt012a.pdf?ts=1724579855731&ref_url=https%253A%252F%252Fitem.szlcsc.com%252F}{DS90LV012ATMF/NOPB}. To achieve the shortest propagation delay, the \href{https://www.ti.com/cn/lit/ds/symlink/tlv3801.pdf?ts=1724579913716&ref_url=https%253A%252F%252Fitem.szlcsc.com%252F}{TLV3801DSGT} LVDS comparator is employed instead of the traditional Transistor-Transistor Logic/Low-Voltage Transistor-Transistor Logic (TTL/LVTTL) discriminator chips; it receives the SiPM signal and generates the LVDS signal. The \href{https://www.ti.com/cn/lit/ds/symlink/ds90lt012a.pdf?ts=1724579855731&ref_url=https%253A%252F%252Fitem.szlcsc.com%252F}{DS90LV012ATMF/NOPB} functions as a converter that translates the LVDS signal into the LVTTL signal. A $50\Omega$ resistor is connected in parallel to the input pin of the \href{https://www.ti.com/cn/lit/ds/symlink/tlv3801.pdf?ts=1724579913716&ref_url=https%253A%252F%252Fitem.szlcsc.com%252F}{TLV3801DSGT} for impedance matching.

\begin{figure}[htbp]
\centering 
\subfigure[A completed PCB board for the gate detector. All the electronic components are shown, including the step-up DC-DC converter, discriminator, and trigger signal output.]{\includegraphics[width=0.9\linewidth, origin=c]{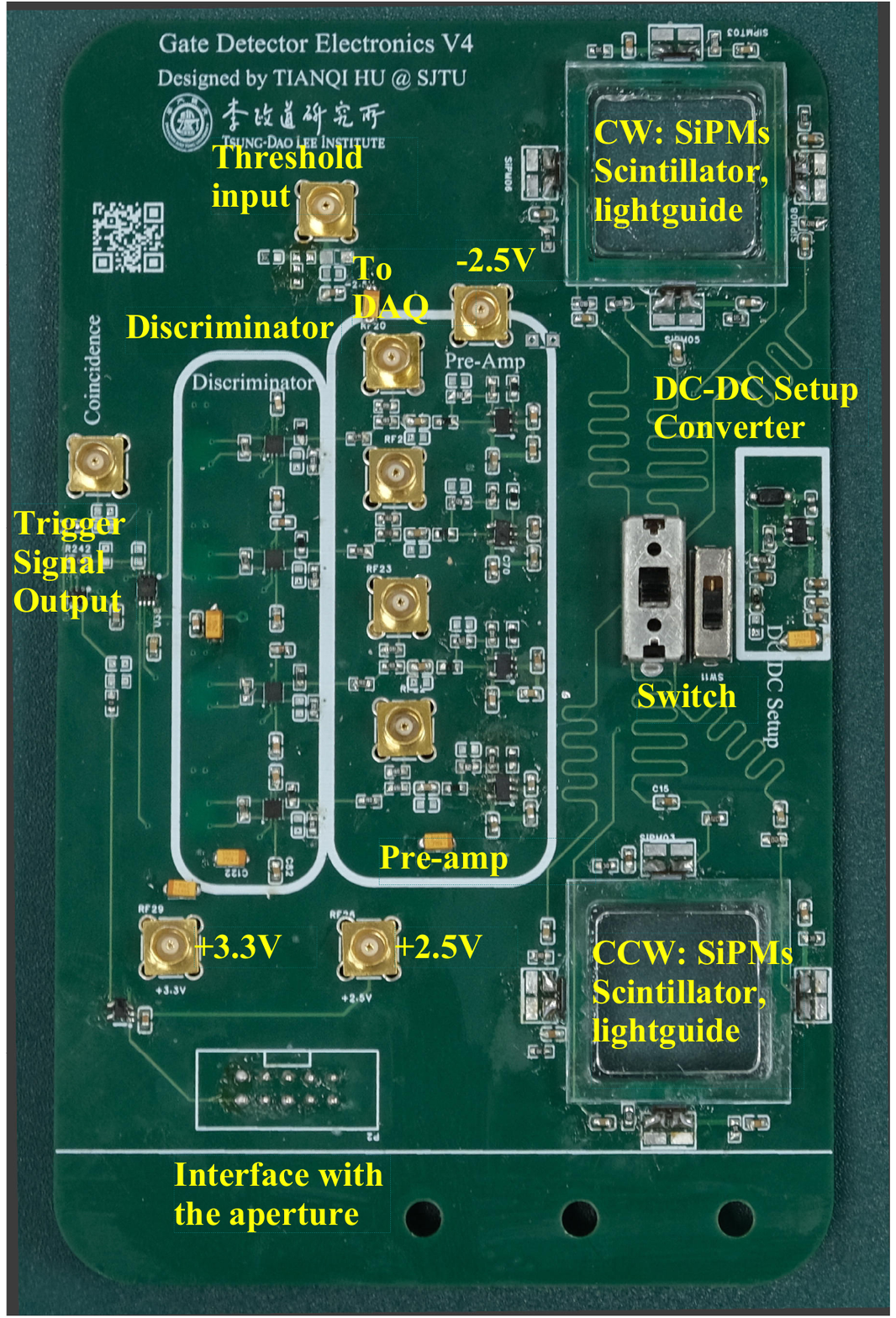}\label{fig:Gate_PCB}}
\qquad
\subfigure[A completed PCB board for the aperture detector. All electronic components are displayed, including the interface with the gate detector, discriminator, and pre-amplifier.]{\includegraphics[width=0.9\linewidth,origin=c]{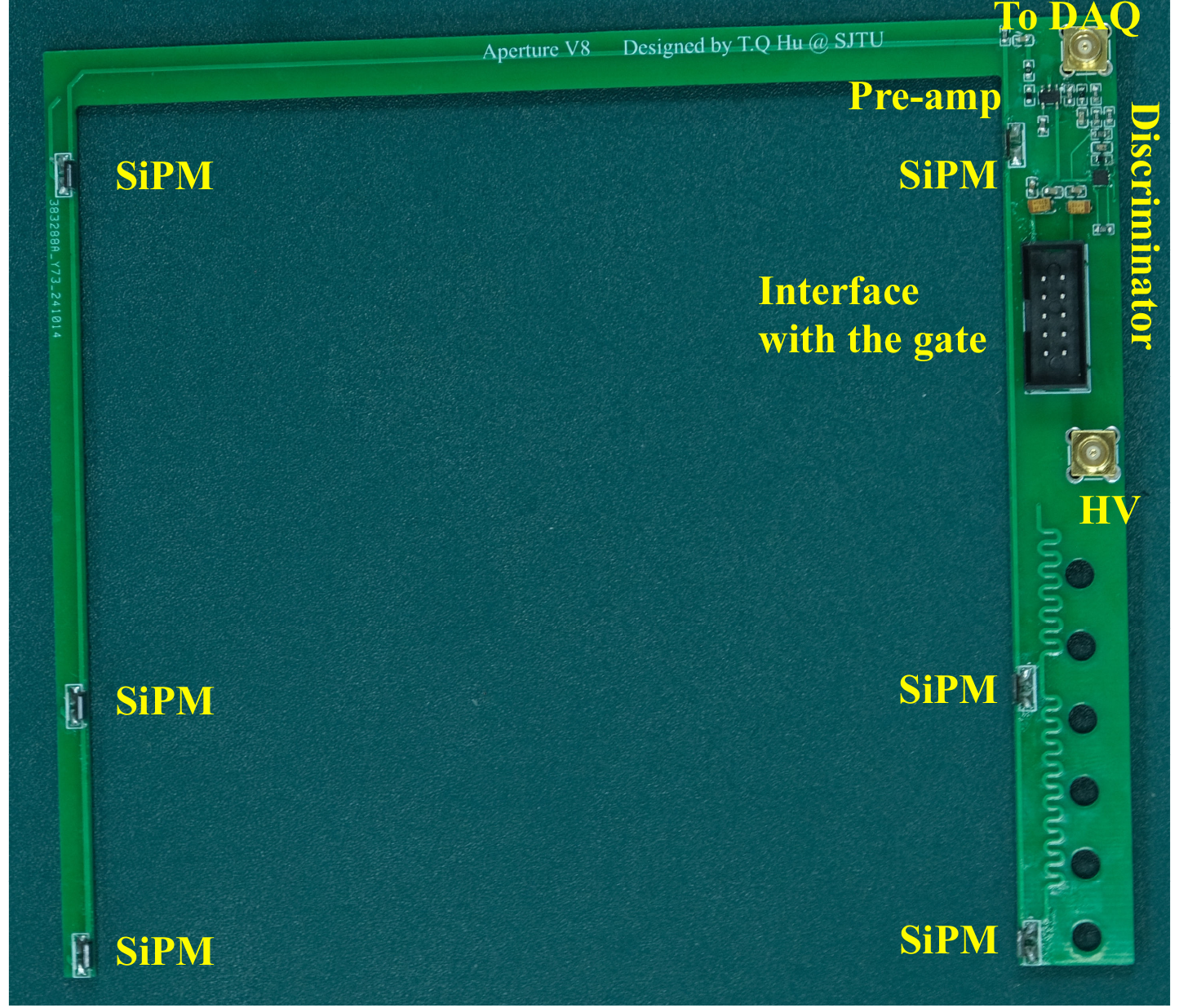}\label{fig:Aperture_PCB}}
\caption{\label{fig:electronics_prototype} Completed PCB boards for the muon trigger detector.}
\end{figure}

\subsection{Coincidence Module}

As illustrated in Fig.~\ref{fig:AntiCoincidence_circuit}, the coincidence circuit comprises four integrated circuit chips. The LVTTL signals from the SiPMs in the gate detector are routed into the AND logic chip, specifically the \href{https://www.ti.com/product/SN74AUC2G08?keyMatch=SN74AUC2G08DCUR&tisearch=universal_search&usecase=OPN}{SN74AUC2G08DCUR} from Texas Instruments. The \href{https://www.ti.com/product/SN74AUC2G08?keyMatch=SN74AUC2G08DCUR&tisearch=universal_search&usecase=OPN}{SN74AUC2G08DCUR} includes two AND gates, each with two inputs. Consequently, the four SiPMs in the gate detector generate two sets of trigger signals, with each pair of SiPMs producing one signal. Similarly, the LVTTL signal from the aperture detector is sent to a NAND logic chip, specifically the \href{https://www.ti.com/product/SN74AUC1G00/part-details/SN74AUC1G00DCKR}{SN74AUC1G00DCKR} from Texas Instruments.

The outputs from these three logic chips are fed into an AND logic chip, the \href{https://www.ti.com/product/SN74LVC1G11}{SN74LVC1G11DCKR} from Texas Instruments. This chip includes one AND gate and three input pins that execute the anti-coincidence operation. This setup enables the coincidence readout of the four SiPMs at the gate detector, effectively determining whether a trigger signal for the HV switch is generated, while the signal from the aperture detector has the authority to block the generation of this signal.

\section{The prototype boards of the electronics}

The prototype boards for the readout electronics have been produced according to the schematic diagrams presented in Fig.~\ref{fig:electronics_prototype} for the gate and aperture electronics boards. The gate board has two windows to accommodate muon injections from Clockwise (CW) and Counter-Clockwise (CCW) directions. Each window features an ultra-thin EJ-200 scintillator slice measuring $20 \times 20 \times 0.1\,\text{mm}^3$, housed within a $25 \times 25 \times 5\,\text{mm}^3$ acrylic light guide. The light guide is fabricated in our laboratory using Computer Numerical Control (CNC) drilling and grinding techniques to improve photon collection. The four Silicon Photomultipliers (SiPMs) are attached to the center of each side of the light guide with BC-603 optical grease, ensuring efficient detection of the collected photons.

\begin{figure}[htbp]
\centering 
\subfigure[The photon spectrum was measured in an electronics lab at PSI using a setup consistent with the beam test. A 2.5 photoelectron threshold was applied, and data were acquired with a DRS4 evaluation board. The integration window spanned from \SI{30}{ns} before to \SI{120}{ns} after the signal peak.]{\includegraphics[width=.45\textwidth, origin=c]{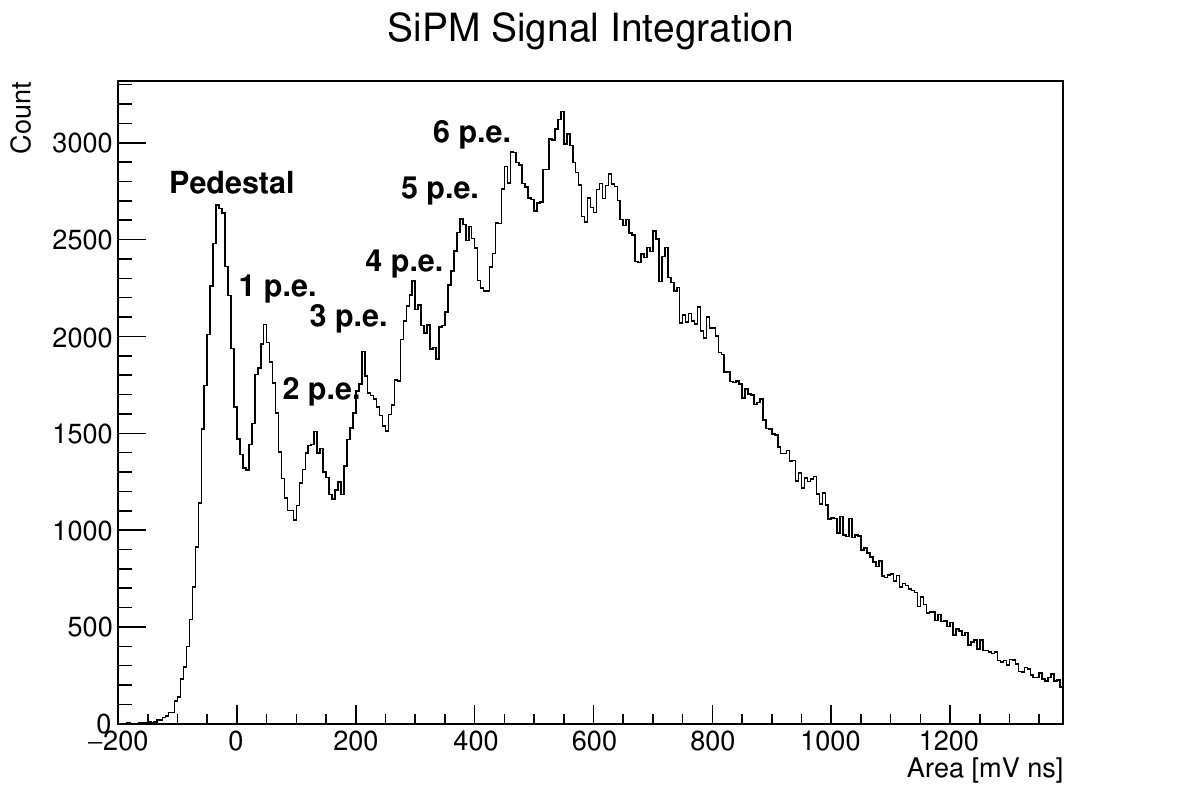}\label{fig:PeakArea}}
\qquad
\subfigure[Photon Response Linearity. Data points were extracted via multi-Gaussian fits to the measured spectra, yielding the mean positions and uncertainties of each photoelectron peak.]{\includegraphics[width=.45\textwidth,origin=c]{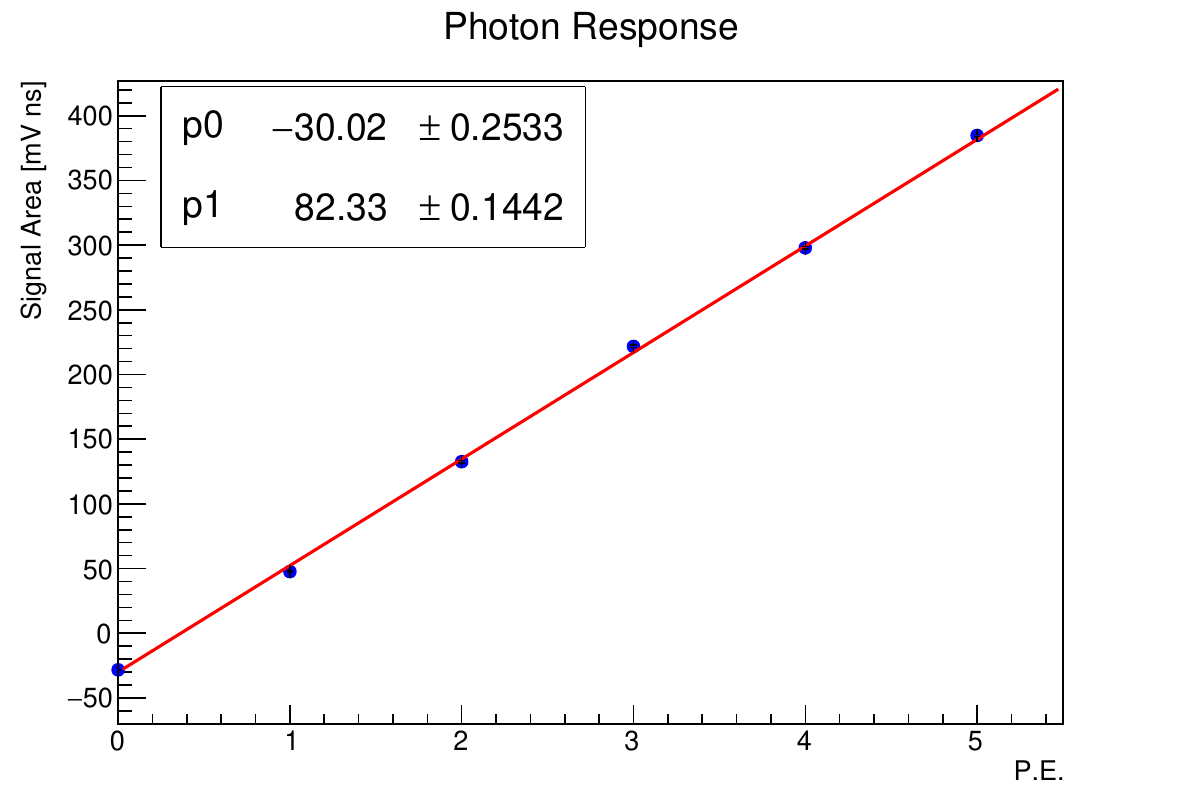}\label{fig:Linearity}}
\caption{\label{fig:SiPMResponseLinearity} Linearity test of the SiPM response after the signal splitter output, performed using an Am-241 radioactive source placed at the gate scintillator slice.}
\end{figure}

Additionally, two switches have been integrated into the SiPM power supply and signal readout circuits on the gate PCB, allowing for the seamless activation and deactivation of the corresponding SiPMs in both CW and CCW modes. This configuration enables the CW and CCW SiPMs to share a set of readout channels, effectively reducing the total number of electronic channels from eight to four. Moreover, at the gate detector, the signal line lengths for the CW and CCW SiPMs have been adjusted to match, ensuring that all signals reach the coincidence circuit simultaneously, unaffected by variations in path lengths. As shown in Fig.~\ref{fig:SiPMResponseLinearity}, the SiPM signal from the splitter output on the gate PCB exhibits a strong linear response in the low photon range, with a fitted slope of 82.33~mV$\cdot$ns/pe between the integrated signal area and the number of photoelectrons (pe). This linear region is particularly relevant, as the gate detector also operates within this low-photon regime, where the number of photons collected by a single SiPM is estimated to be between 9 and 20, according to Geant4 optical photon simulations.

The amplitude of a single photoelectron (SPE) was determined by analyzing the photoelectron spectrum at different threshold settings. At a threshold of \SI{3}{mV}, the SPE peak was clearly visible, while at \SI{5}{mV}, it disappeared. Based on this and tests with higher photoelectron peaks, the amplitude of a single photoelectron was estimated to be approximately \SI{4}{mV}. This estimation was confirmed by monitoring the signal over time, where both the amplitude and shape of the SPE peak remained stable and reproducible.

The size and shape of the aperture electronics board are designed according to the dimensions of the scintillator ($130 \times 135 \times 5\,\text{mm}^3$), with the six SiPMs positioned where the maximum photon count occurs at the edge, based on simulation results. The six SiPMs are connected in parallel to a single output, mainly to minimize the number of channels required, as their primary function is to provide trigger signals. This configuration results in a six-fold reduction in internal resistance and a six-fold increase in internal capacitance. Consequently, the signal amplitude decreases, and the rise time increases, with the amplitude of a single photoelectron corresponding to \SI{0.67}{mV}. However, this arrangement is sufficient for the intended purpose of generating trigger signals. Communication with the gate detector is established through ribbon cables, while the anti-coincidence circuit is embedded in the gate detector's circuit board to manage both coincidence and anticoincidence operations.

\section{Beam Test at PSI $\pi$E1 and results}

\begin{figure}[htbp]
\centering 
\includegraphics[width=.5\textwidth, origin=c]{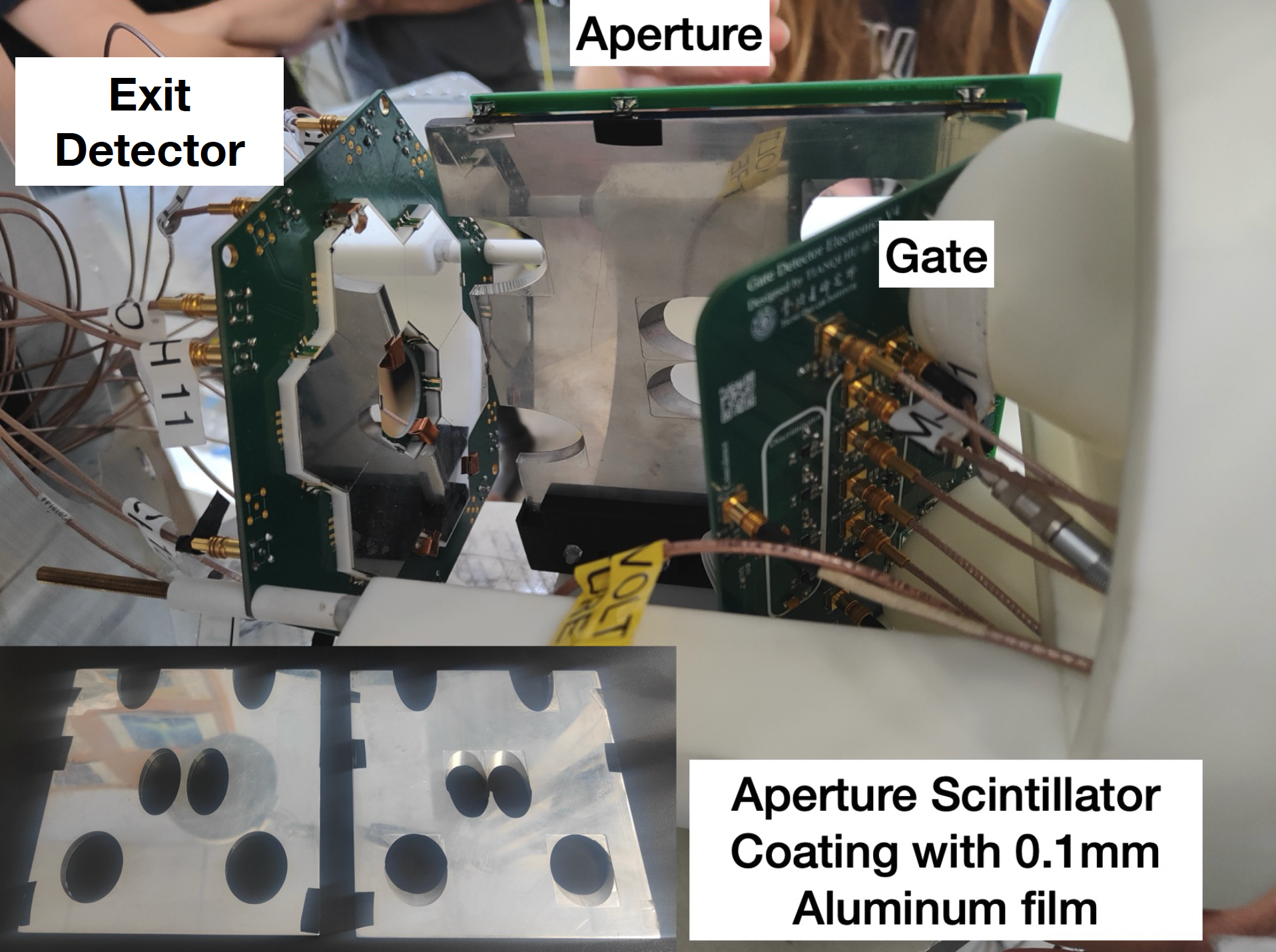}
\caption{Muon trigger detector prototype for the Beam Test 2024 at the PSI $\pi E1$ beamline.}
\label{fig:Beam2024_layout}
\end{figure}

From October 28 to November 5, 2024, the muon trigger detector prototype, along with its fast electronics readout circuit, underwent a week-long beam test at the PSI $\pi E1$ beamline, where tests were conducted using \SI{22.5}{MeV/c} muons and \SI{7}{MeV/c} positrons. Figure~\ref{fig:Beam2024_layout} illustrates the setup of the muon trigger detector, highlighting how the exit detector records the particles' tracks after they pass through it. The aperture scintillator was coated with \SI{0.1}{mm} thick aluminum foil to prevent optical crosstalk between the gate scintillator and the aperture scintillator. The detectors are located at the bore of the solenoid, which generates a uniform magnetic field of \SI{780}{mT}. Magnetic steel tubes were installed for this beam test instead of the superconducting injection channel intended for Phase I. The upstream beam intensity was $4 \times 10^{6}$\,Hz, whereas the beam rate prior to entering the magnetic steel tube was $2 \times 10^5$\,Hz for the $\mu^{+}$ and $4 \times 10^3$\,Hz for the $e^{+}$.

The positron and muon were emitted from the beamline, first passing through a \SI{20}{cm} air gap, then traveling through the injection tube, which consists of \SI{800}{mm}-long tubes with an inner diameter of \SI{15}{mm} that define the initial collimation, before reaching the muon trigger detector. Upon arrival at the muon trigger detector, the positrons had a momentum of \SI{7}{MeV/c}, while the muons had a momentum of \SI{22.5}{MeV/c}. Particles within the acceptance range of the muon trigger detector pass through the holes in the aperture scintillator and continue to the exit detector. In contrast, particles outside this range are stopped and absorbed by the aperture scintillator. The rate emitted from the tubes is approximately \SI{270}{Hz}.

\begin{figure}[htbp]
\centering 
\subfigure[Wavefrom for Anti-Coincidence Trigger. The $e^{+}$ hits only the gate detector and results in the trigger signal output.]{\includegraphics[width=.46\textwidth, origin=c]{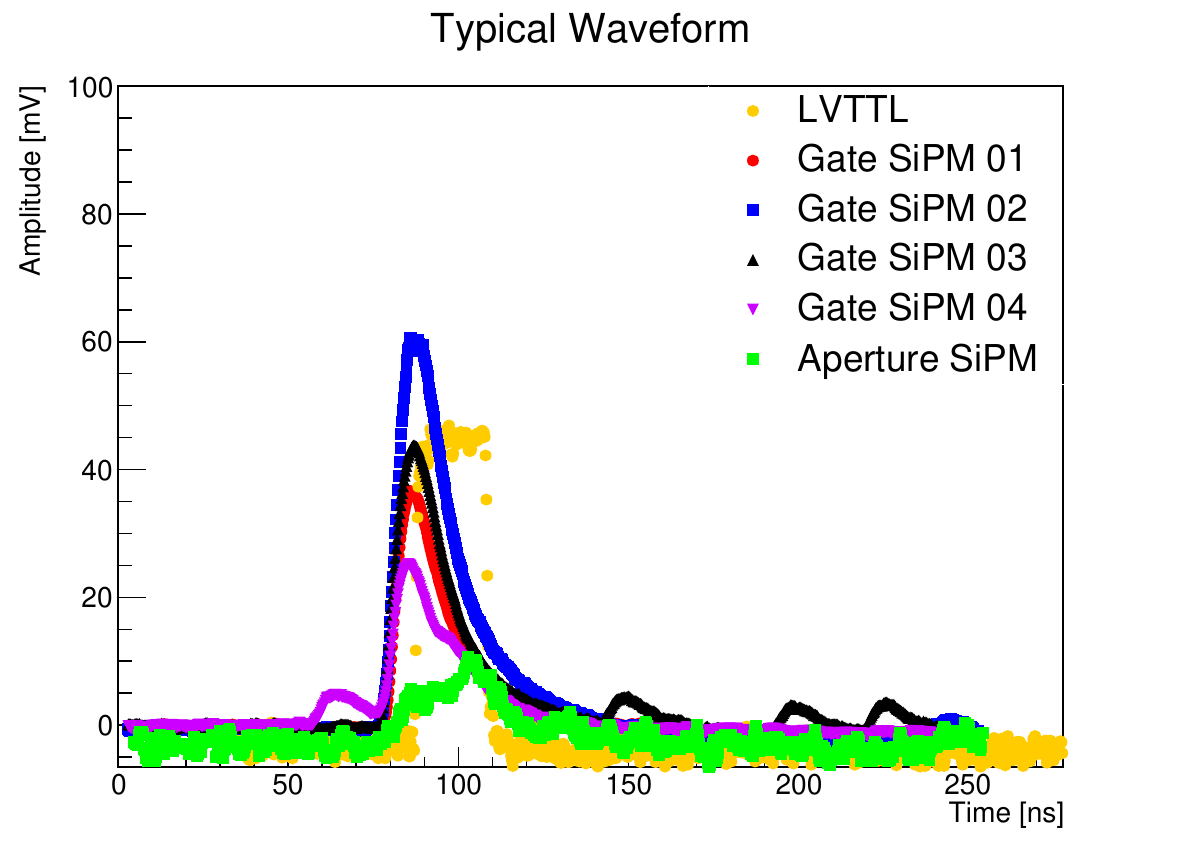}\label{fig:AntiCoincidence_Beam2024}}
\qquad
\subfigure[Wavefrom for Coincidence Trigger. The $e^+$ hits both the gate and the aperture detector, in which the signal from the aperture detector prevents the production of the trigger signal.]{\includegraphics[width=.46\textwidth, origin=c]{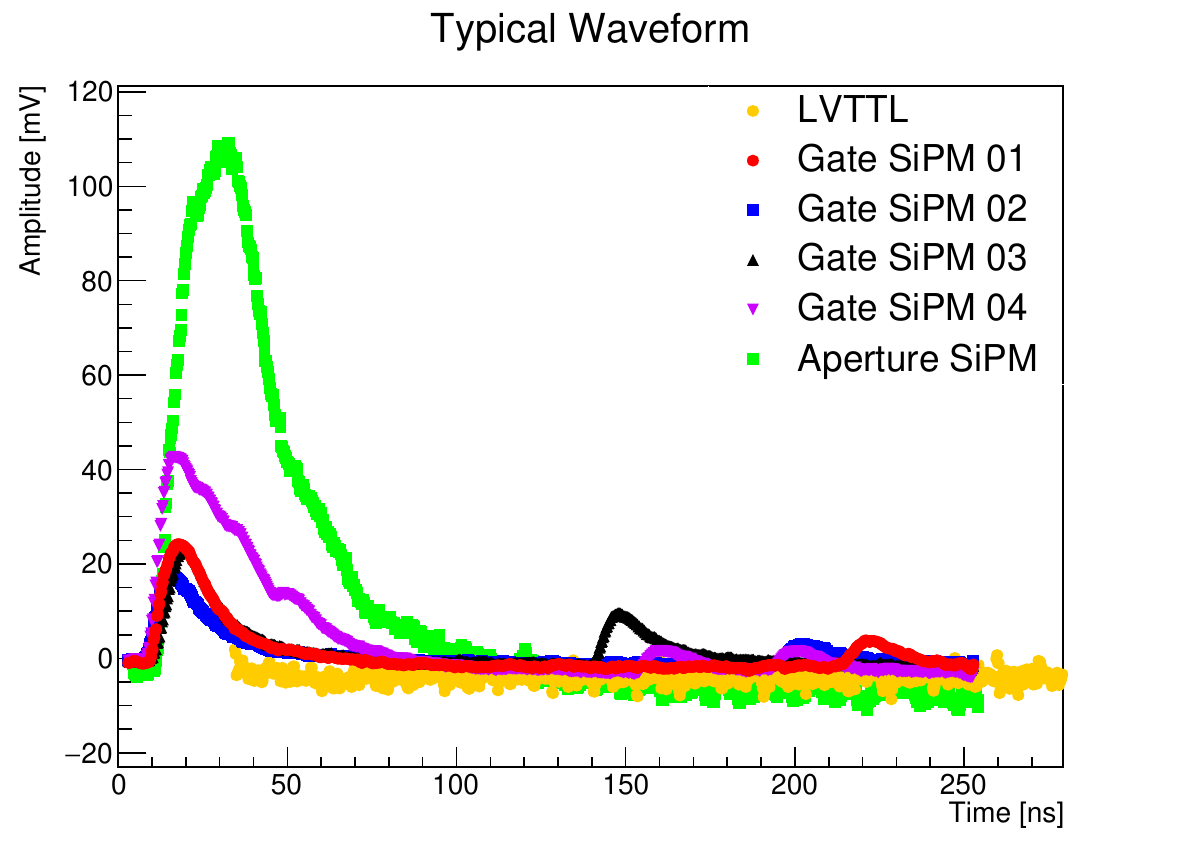}\label{fig:Coincidence_Beam2024}}
\caption{\label{fig:Typical_Waveform_Beam2024} Typical waveform recorded by the WaveDREAM board for the muon trigger detector under positron injection. The LVTTL trigger signal, which is used to activate the pulsed magnetic field, is transmitted through a cable equipped with a 32\,dB attenuator, in order to reduce its amplitude to within the dynamic range of the WaveDAQ system.}
\end{figure}

The WaveDAQ system, originally developed for the MEG II experiment~\cite{Francesconi:2023cxt}, is employed for waveform digitization in this setup. It utilizes DRS4 chips operating at a sampling rate of 5~GSPS, with each triggered event yielding 1024 digitized samples. The system features a dynamic range of $\pm$\SI{0.5}{V} and achieves a signal-to-noise ratio equivalent to 11.5 effective bits, enabling accurate reconstruction of fast analog signals with high time and amplitude resolution. The standard waveforms corresponding to coincidence (when both the gate and aperture detectors are activated) and anticoincidence (when only the gate detectors are activated) are presented in Fig.~\ref{fig:Typical_Waveform_Beam2024}, with coincidence and anti-coincidence rates of approximately \SI{250}{Hz} and \SI{10}{Hz}, respectively.

\begin{figure}[htbp]
\centering 
\includegraphics[width=.46\textwidth, origin=c]{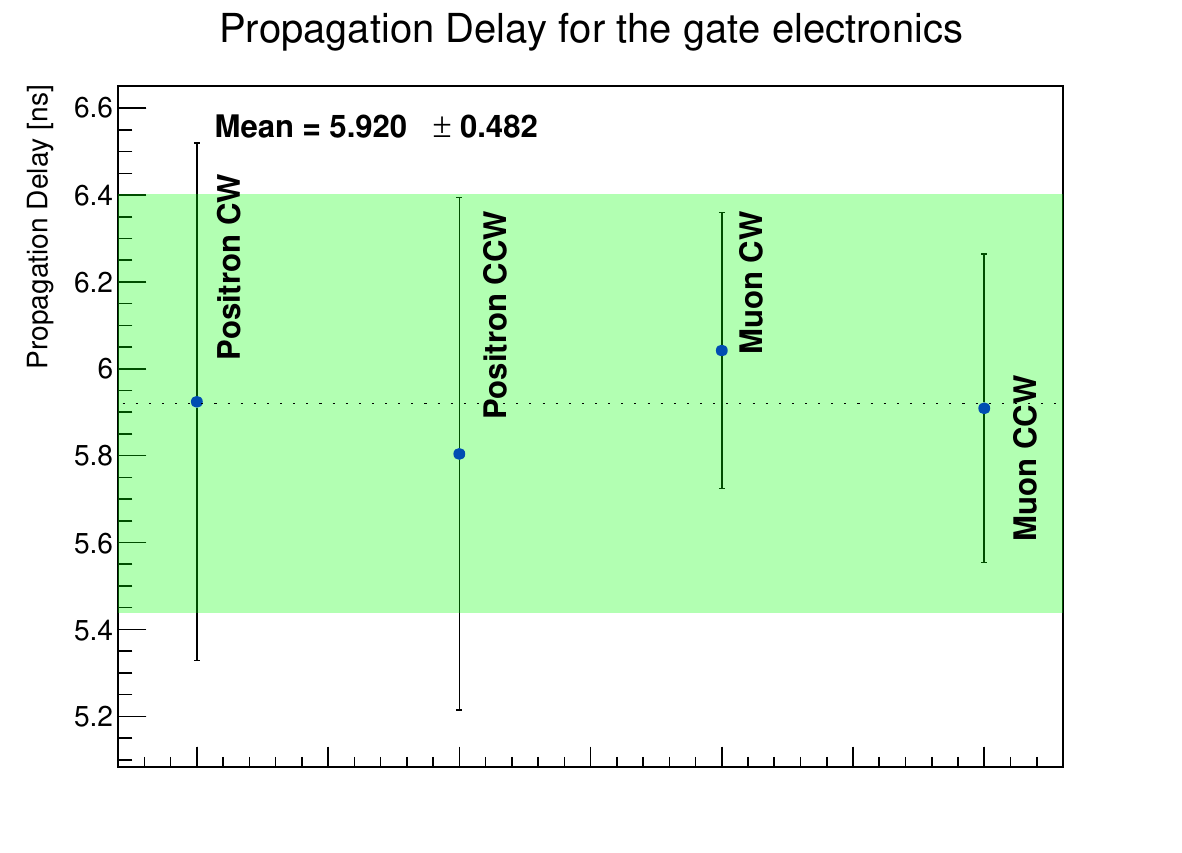}
\caption{Propagation delay results for the Beam test 2024.}
\label{fig:Propagation_Delay_Beam2024}
\end{figure}

The propagation delay is defined as the time interval during which all SiPM signals exceed the threshold (\SI{10}{mV} for $e^+$, \SI{15}{mV} for $\mu^+$) to the half-height of the rising edge of the LVTTL trigger signal output. The results of the propagation delay for both CW and CCW muons, as well as positrons, are presented in Fig. ~\ref{fig:Propagation_Delay_Beam2024}, with an average of $5.92(0.48)$ \SI{}{ns}. More importantly, the electronics remained stable under the magnetic field environment and during long-term operation, as indicated by the datasets of the CW/CCW $\mu^+$/$e^+$ at different time points, showing no significant discrepancies in the propagation delay measurements. The error bar for the $\mu^+$ is smaller than that for the $e^+$ for two main reasons: first, the statistics for the $\mu^+$ are ten times greater than those for positrons; second, the DAQ settings and beam tuning for the $\mu^+$ were optimized compared to those for $e^+$.

\begin{figure}[htbp]
\centering 
\includegraphics[width=.46\textwidth, origin=c]{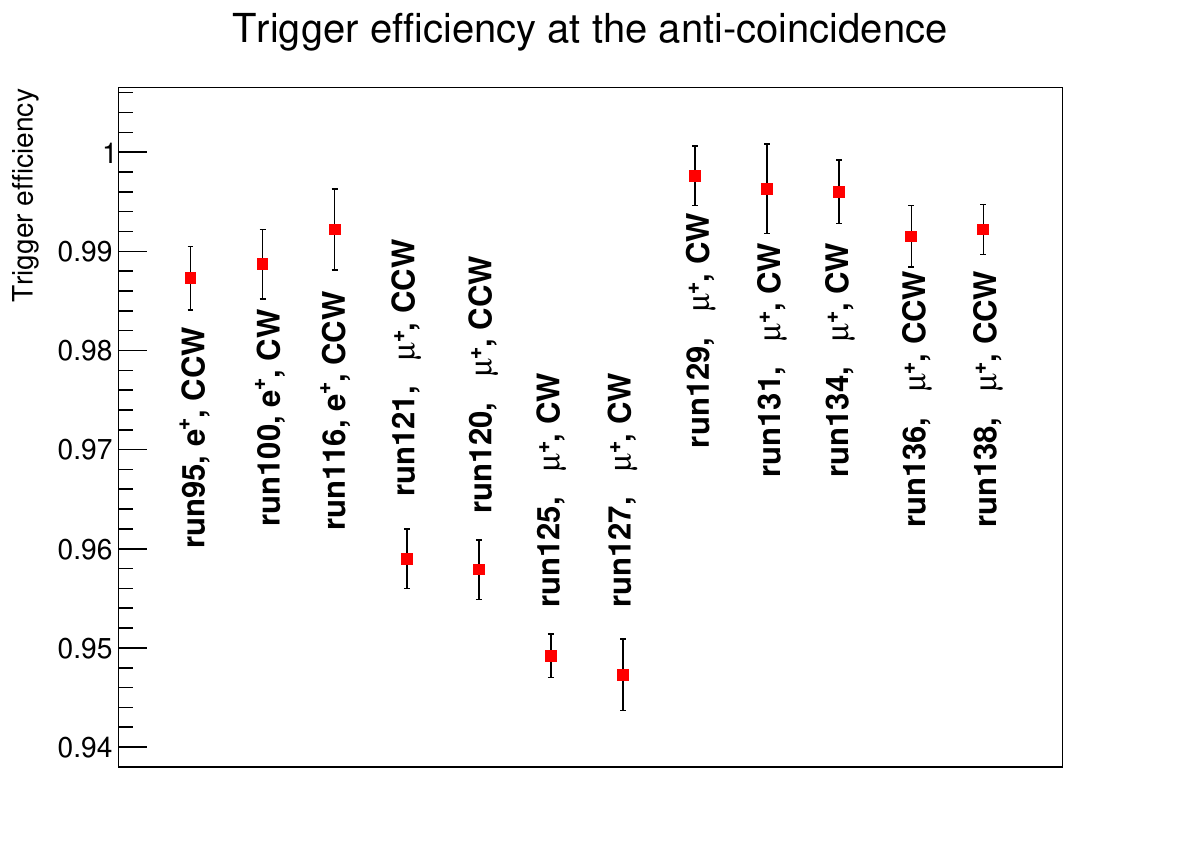}
\caption{Anti-coincidence efficiency for the $\mu^{+}$ and $e^{+}$ datasets under various conditions. It should be noted that for datasets run\,120, run\,121, run\,125, and run\,127, the separator of the beamline was found to be turned off, as most of the $\mu^{+}$ events in these datasets are contaminated by the $e^{+}$.}
\label{fig:Anticoincidence_efficiency}
\end{figure}

The anti-coincidence efficiency of the detector is defined as the effectiveness of the electronics in outputting LVTTL trigger signals under anticoincidence conditions. When a particle strikes the gate detector and successfully passes through the holes of the aperture detector, a trigger signal should be generated, as illustrated in Fig.~\ref{fig:AntiCoincidence_Beam2024}. Figure~\ref{fig:Anticoincidence_efficiency} shows the distribution of the anti-coincidence efficiency for the datasets. The anti-coincidence efficiency for $\mu^+$ exceeds 99\%, while for $e^{+}$ it exceeds 98\% (the runs 120, 121, 125, and 127 were not used for the $\mu^{+}$ anti-coincidence efficiency calculation. This is because the separator was not activated during these runs, resulting in significant contamination from secondary particles). The reasons for the discrepancy are twofold: 

\begin{enumerate}
    \item The hysteresis of the discriminator requires the signal amplitude to exceed a threshold value of approximately \SI{2}{mV} to activate the discriminator. For instance, with a \SI{10}{mV} threshold, the discriminator activates only when the signal exceeds about \SI{12}{mV}. This results in a 1-2\% ratio of anticoincidence cases where, even though the signal amplitude surpasses the discriminator threshold, it does not overcome the hysteresis, thus preventing the trigger signal from being output. This effect will be considered when setting the threshold for collecting the physics data.   
    \item The energy deposition within the gate detector is greater for the positively charged muons ($\mu^+$) than for the positrons ($e^+$), with average energy depositions recorded as $\bar{E}_{\mu^{+}} = 0.272$ MeV and $\bar{E}_{e^{+}} = 0.017$ MeV, respectively. As a result, the signals generated by the silicon photomultipliers (SiPM) from the $\mu^+$ exhibit a larger amplitude, facilitating their ability to exceed the threshold and surpass the hysteresis of the discriminator.
\end{enumerate}
 
The efficiency of coincidence is defined as the probability that the aperture detector prevents the output of LVTTL trigger signals under conditions of coincidence. When a particle strikes both the gate and the aperture detector, the signal generated by the aperture detector should inhibit the production of the trigger signal, as depicted in Fig.~\ref{fig:Coincidence_Beam2024}. In the context of coincidence events, 97(5)\% did not result in the output of the LVTTL trigger signal, whereas 2.6(8)\% of coincidence events did generate the LVTTL trigger signal. Although these unexpected LVTTL trigger signals may activate the pulsed magnetic field, no muons are emitted, as they remain confined within the aperture scintillator. Consequently, the data acquisition system records the events as empty. Assuming that 95\% of the muons successfully enter the storage area, the proportion of empty events would amount to less than 0.1\% of the total events. Should the success rate decrease to 90\%, this proportion would increase to less than 0.3\%, and at a success rate of 80\%, it would be less than 0.6\%. These empty events may be excluded through offline analysis or alternative methods.

The failure of coincidence primarily arises from the substantial capacitance resulting from the parallel configuration of six SiPMs in the aperture detector, which leads to an extended rise time of the SiPM signal. As a consequence, approximately 2.62(0.83)\%  of the coincidence events possess time windows for the trigger signal from the aperture detector that do not completely coincide with those of the gate trigger signal, resulting in the generation of a narrow LVTTL trigger width.

The electronics properties are detailed in Tab.~\ref{tab: properties_of_the_electronics}, meeting the requirements of the PSI muEDM experiment. The next steps will focus on further optimizations, such as refining the wiring of the SiPMs at the aperture detector, adjusting the connections and cable lengths between the gate and aperture detectors, adding a temperature feedback module, and more. These enhancements will establish the foundation for the upcoming beam test in 2025.

\begin{table}[!htb]
\centering
\caption{\textsc{Properties} of the \textsc{Muon Trigger Detector Prototype} and \textsc{PSI muEDM Requirements}.}
\label{tab: properties_of_the_electronics}
\smallskip
\def\arraystretch{1.2}
\begin{tabular}{|c|c|c|}
\hline
& Electronics Board & muEDM requirements\\ \hline
Anti-Coincidence efficiency  & 99.5\% & $\geqslant$ 95\%\\ \hline
Coincidence efficiency & 97.4\% & $\geqslant$ 95\%\\ \hline
Propagation delay [ns] & $5.9 \pm 0.5$ & $\leqslant$ 15\\ \hline
Trigger Signal Voltage [V] & 2.5 & $>$ 1.7\\ \hline
Trigger Signal Width [ns] & 20.0 $\pm$ 6.0 & $>$ 3\\ \hline
\end{tabular}
\end{table}

\section{Conclusion}

In this work, we presented the development and performance evaluation of the fast front-end electronics for the muon trigger detector in the PSI muEDM experiment. The system was designed to meet stringent requirements, including minimal signal propagation delays, high detection efficiency, and stable operation under high-magnetic-field conditions. To achieve these objectives, we implemented a custom-designed PCB layout with non-magnetic components, optimized readout circuits, and an anti-coincidence trigger logic to efficiently identify storable muons while rejecting non-storable events.

The beam test conducted at the PSI $\pi$E1 beamline validated the performance of the trigger detector and its electronics. The results demonstrated:
\begin{itemize}
    \item High anti-coincidence efficiency of 99.5\% for muons, exceeding the required threshold of 95\%;
    \item Coincidence efficiency of 97.4\%, ensuring accurate muon identification;
    \item Low propagation delay of $5.9 \pm 0.5$\,ns, well within the experimental requirement of $\leq 15$\,ns;
    \item Stable operation in the \SI{780}{mT} solenoidal magnetic field demonstrates the robustness of the design and lays the foundation for future experimental phases under the \SI{3}{T} field.
\end{itemize}
The system demonstrated excellent timing precision and operational reliability during extended beamline runs, confirming its suitability for long-term data acquisition in the muEDM experiment. The integration of optimized trigger electronics and low-latency signal processing significantly reduced background contamination and enhanced the selection efficiency of storable muons for the frozen-spin technique.

Looking ahead, the system will undergo further upgrades and beam validations to ensure robust performance under the most demanding experimental conditions. Planned improvements include wiring refinements, mitigation of capacitance effects in the aperture detector, and the integration of temperature feedback mechanisms to improve long-term stability. These enhancements will ensure the readiness of the trigger detector system for Phase I of the PSI muEDM experiment, supporting high-precision measurements in the ongoing search for a muon electric dipole moment.




\section*{References}

\def\refname{\vadjust{\vspace*{-1em}}} 

\end{document}